# Ultra-Low Thermal Conductivity in Organic-Inorganic Hybrid Perovskite $CH_3NH_3PbI_3$

*Andrea Pisoni[1]\*, Jaćim Jaćimović[1], Osor S. Barišić[2], Massimo Spina[1], Richard Gaál[1], László Forró[1], Endre Horváth[1].*

[1] Laboratory of Physics of Complex Matter, EPFL, CH-1015 Lausanne, Switzerland

[2] Institute of Physics, Bijenička c. 46, HR-10000 Zagreb, Croatia

AUTHOR INFORMATION

**Corresponding Author**

\* andrea.pisoni@epfl.ch

.

ABSTRACT

We report on the temperature dependence of thermal conductivity of single crystalline and polycrystalline organometallic perovskite $CH_3NH_3PbI_3$. The comparable absolute values and temperature dependence of the two sample's morphologies indicate the minor role of the grain boundaries on the heat transport. Theoretical modelling demonstrates the importance of the resonant scattering in both specimens. The interaction between phonon waves and rotational degrees of freedom of $CH_3 NH_3^+$ sub-lattice emerges as the dominant mechanism for attenuation of heat transport and for ultralow thermal conductivity of 0.5 W/(Km) at room temperature.

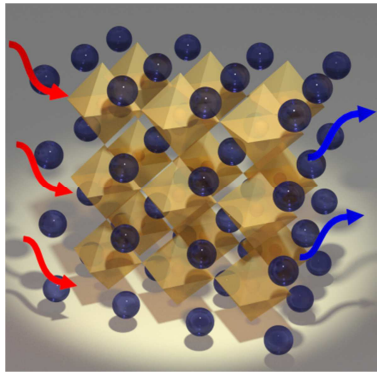

Recently the organic-inorganic hybrid compound $CH_3NH_3PbI_3$ (hereafter $MAPbI_3$) has emerged as the central component of highly efficient solar cells[1-4]. $MAPbI_3$ is deposited in a dye sensitized solar cell configuration by spin coating on $TiO_2$ nanoparticles and using a solid state redox mediator its light conversion efficiency ($\eta$) reaches 16%, an unprecedented value for such a device. There is very vivid world-wide activity to understand the success of this material in solar cells and to find the materials parameters which would be tuned to further improve $\eta$ [5-6]. The known advantages of this material are the high cross section for photo-electron generation, the long diffusion length of the charge carriers and the simplicity of $MAPbI_3$ synthesis and device architecture [4]. The unknown parameters are the health hazards related to the handling of $MAPbI_3$ and the expected lifetime of these solar cells.

The thermal management of such multi-component solar cells can be an important factor in the device's lifetime since a large part of the solar radiation is converted into heat [7]. The lack of its evacuation can result in mechanical stresses in the sandwich-structured device and over time it can cause structural decoupling of the constituents and severely decrease $\eta$. Thus the heat distribution in the cell and particularly around $MAPbI_3$ must be controlled.

The scope of this paper is therefore to study and understand the thermal transport occurring in $MAPbI_3$ as a necessary pre-requisite to well-designed new devices. To our knowledge no investigation of thermal conductivity ($\kappa$) of $MAPbI_3$ has been performed so far. Moreover, since thermal conductivity is strongly dependent on the sample's morphology [8-9] we analyze the change in $\kappa$ in single crystal and in polycrystalline $MAPbI_3$ samples. The measurements were performed on few $mm^3$ large single crystals (SC) and on polycrystalline sample (PC) obtained by pressing together an assembly of micro-crystallites. PC mimics the material's texture in solar cell devices. Typical samples and the zoom on the microstructure are shown in figures 1 and 2 for single and polycrystalline samples, respectively. Resistivity measurements confirm the insulating nature of the material. Regarding thermal properties we find ultralow values of thermal conductivity even on single crystals which is ascribed to the complex unit cell and the disordered $CH_3NH_3^+$ sublattice [10]. Its temperature dependence is described by the *Callaway* formalism arguing that phonons are the main heat carriers in both single crystal and polycrystalline samples.

$MAPbI_3$ crystals were prepared by precipitation from a concentrated aqueous solution of hydroiodic acid containing lead (II) acetate and a respective amount of $CH_3NH_3^+$ solution. The two ends of the sample holder were held at 55 and 42 °C respectively to induce the saturation of the solute at the low temperature part of the solution. After 24 hours sub-millimetre sized crystals appeared in the solution. Large $MAPbI_3$ crystals with 3x5 mm silver-grey mirror-like facets were grown after 7 days. Smaller crystals were ground in a mortar and pressed together to obtain a mechanically stable polycrystalline sample.

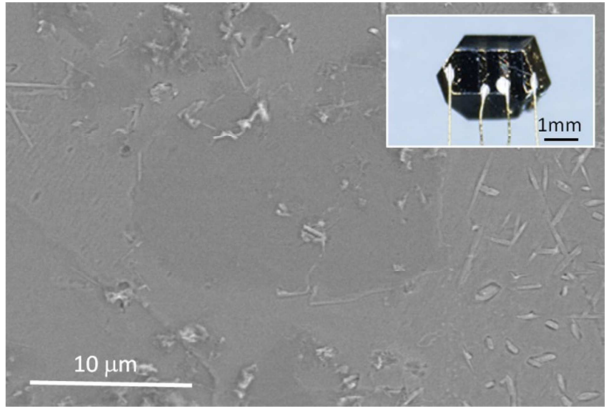 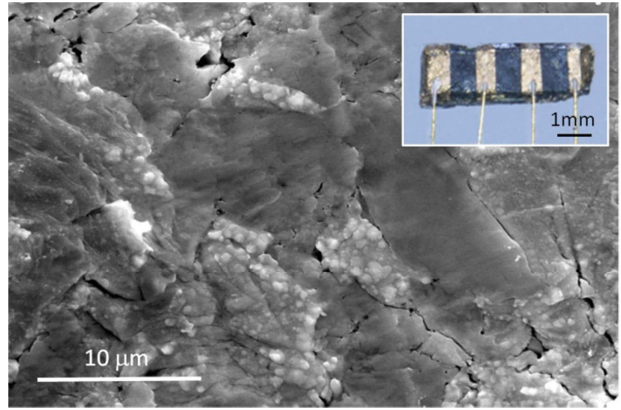

**Figure 1.** SEM image of a CH$_3$NH$_3$PbI$_3$ single crystal. The inset shows a typical crystal used in electrical and thermal transport measurements.

**Figure 2.** SEM image of a CH$_3$NH$_3$PbI$_3$ polycrystalline sample. The inset shows a typical polycrystal used in electrical and thermal transport measurements.

Figure 3 shows the resistivity as a function of temperature for the single crystal and polycrystalline samples measured in the dark. Room temperature resistivity of $13*10^6$ Ωcm in MAPbI$_3$ single SC is in agreement with previously reported data [11]. Upon cooling and heating in the 4-300 K range the sample shows hysteretic behaviour as it has been already observed by Stoumpos et al. [11]. They explain it as a result of the structural phase transition occurring at 162 K. Optical microscopy observations reveal that the change of the lattice parameters at low temperature causes formation of micro-cracks which develop with thermal cycling. Modelling resistivity in the 250-320 K temperature range with thermally activated behaviour $\rho_0 \exp(E_i/k_B T)$ yields an activation energy of $E_i$=185 meV which is ascribed to simple thermal activation from impurity levels (the optical gap is 1.5eV [5]). A similar fit below 250 K gives $E_i$= 70 meV which is likely to come from hopping conduction within the impurity level. Such change in the conductivity mechanism is well known and was reported in different insulating materials like TiO$_2$ [12], ZnO etc.

For PC samples $\rho$(300K)= 38 MΩcm (Figure 4) that is less than a factor of two higher than that of single crystals. The resulting activation energies in the same temperature ranges are 427 meV and 81 meV. These higher values, compared to SC, are interpreted as a consequence of barriers for charge transport at the grain boundaries.

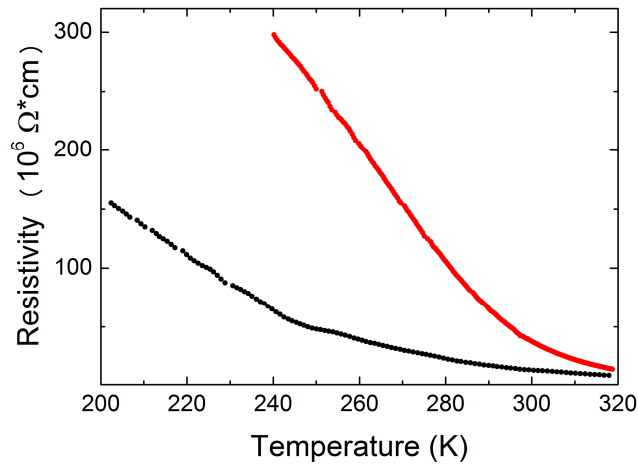

**Figure 3.** Resistivity as function of temperature for single (black curve) and polycrystalline (red curve) samples. In both cases the activated behavior comes from an impurity level at Ei= 0.18 and 0.48 eV from the band edge, respectively.

Figure 4 shows our main result, the temperature dependence of the thermal conductivity of SC and PC MAPbI$_3$. The experiments were repeated on three different specimens with high reproducibility proving that the observed behaviour is an intrinsic material property. The overall temperature dependence of the SC and PC samples is similar. Yet, certain differences are manifested and will be discussed in more detail below.

The room temperature values are 0.5 and 0.3 W/(mK) respectively. These are considered as ultra-low and markedly different from the known inorganic materials such as: $TiO_2$ [13], $Bi_2Te_3$ etc which are in the range of 10-100 W/mK [14]. In fact, the thermal conductivity of $MAPbI_3$ is closer to that of polymeric materials than those of crystalline structures [15].

In the temperature dependence a sharp dip in κ emerges around 160 K. This coincides with the tetragonal - orthorhombic structural phase transition, being previously observed by heat capacity measurements [16]. As shown in the inset of Figure 4, for the SC samples the width of this dip is only 2 K, while for the PC samples the structural transition is smeared over a wider temperature range, corresponding to approximately 8 K.

Apart this phase transition that does not influence much the thermal conductivity, we observe that the temperature dependence resembles that of insulating inorganic materials [17], where three temperature regimes may be distinguished. In the high-temperature regime κ increases with decreasing temperature, and there is a rapid drop in the low-temperature limit $T \rightarrow 0$. Between these two limiting behaviours, κ reaches a maximum, specifically around $T_{max}$ = 30 K and $T_{max}$ = 43 K for the SC and the PC samples respectively.

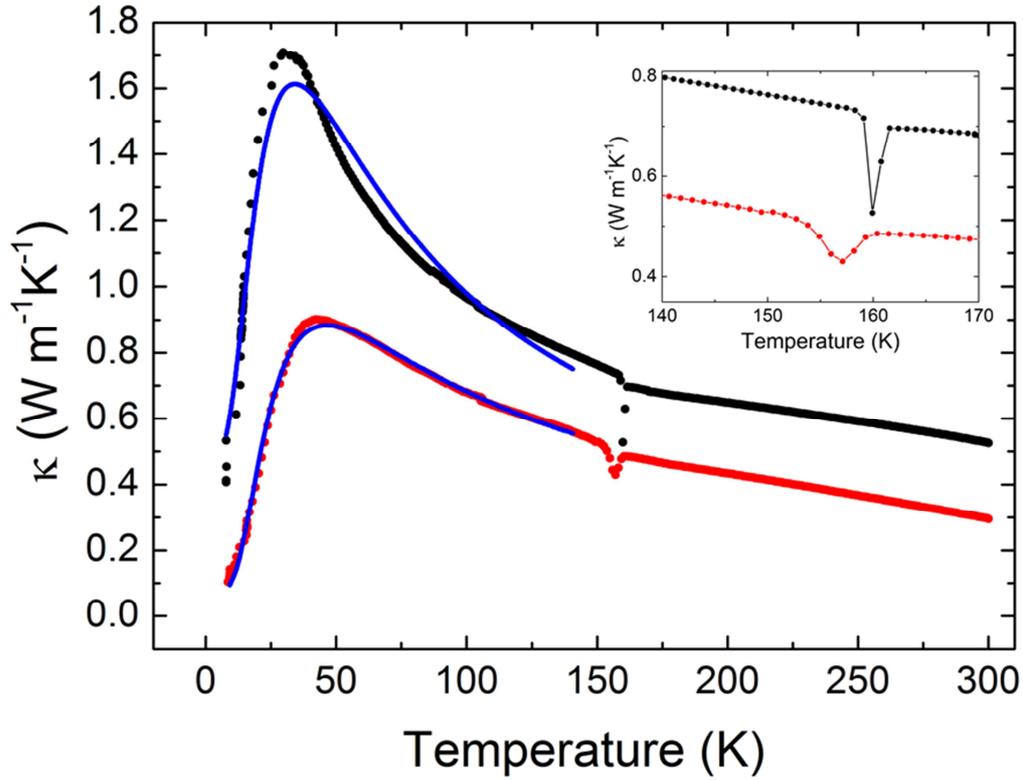

**Figure 4.** Temperature dependence of thermal conductivity of single crystal (black) and polycrystal (red) $CH_3NH_3PbI_3$ samples. Blue lines are obtained from the theoretical model. The inset shows the detail around the structural transition at 160 K.

In an attempt to shed more light on microscopic mechanisms that limit heat conduction, and in particular to resolve differences between the SC and the PC samples, we start our analysis by decoupling the electronic ($\kappa_{el}$) and the lattice ($\kappa_{lat}$) contributions to the total thermal conductivity $\kappa$, since dynamics of these two subsystem should be characterized by very different time scales. For a rough estimate of the former, one may consider the Wiedemann–Franz law $\kappa_{el} = L_0 \sigma T$, where $L_0$ is the Lorenz number and $\sigma$ is the electrical conductivity. In materials for which itinerant electrons contribute considerably to the heat transfer, the order of magnitude of $L_0$ is expected to be around $10^{-8} (V/K)^2$. On the other hand, we find for our MAPbI$_3$ samples that $L_0$ should be four orders of magnitude higher in order to explain thermal conductivity by itinerant electrons only, which clearly implies that these have a negligible role in the total thermal conductivity. Consequently, the lattice contribution κ$_{lat}$ dominates in the entire temperature range.

The temperature dependence of κ$_{lat}$ is frequently analysed within Callaway's approach, which has its origin in the Boltzman equation [18]. The basic approximation is that different phonon scattering mechanism can be treated independently, $\tau^{-1} = \sum_i \tau_i^{-1}$,

$$\kappa = CT^3 \int_0^{\theta_D/T} \tau(x) \left[\frac{x^4 e^x}{(e^x - 1)^2}\right] dx \qquad (1)$$

with $C = \left(\frac{k_B}{2\pi^2 v_s}\right)\left(\frac{k_B}{\hbar}\right)^3$, $x = \hbar\omega/k_B T$, $\theta_D$ Debye temperature, $k_b$ Boltzmann constant, $\hbar$ Planck constant, and $v_s$ the average speed of sound.

Our first approach for the analysis of the data presented in Figure 4 was the traditional one, considering three scattering mechanisms, $\tau^{-1} = \tau_b^{-1} + \tau_p^{-1} + \tau_u^{-1}$, 13. The first one, due to sample boundaries, is proportional to the square root of the inverse cross section of the specimen, $\tau_b^{-1}(x) \propto a_1 \propto 1/S^{1/2}$. The second is associated with scattering events by point defects like vacancies, substitutions, and other point-like impurities. Within the Rayleigh scattering model, the associated relaxation rate is taken to be temperature independent, $\tau_p^{-1}(x) \propto \omega^4 \propto a_2 T^4 x^4$. Finally, as temperature increases, one should consider interactions between the phonons, since with leaving the harmonic regime, three-phonon Umklapp processes start to dominate the behaviour of κ. For the corresponding relaxation rate we use the form $\tau_u^{-1}(x) \propto a_3 T^3 x^2 \exp(-\frac{\theta_D}{\alpha T})$. In the high temperature limit $\theta_D \ll \alpha T$, this form reduces to the appropriate one for describing Umklapp and normal processes $\tau_u^{-1} \propto T^3 x^2$, with κ obeying the 1/T-law [19]. Combining all three relaxation times we obtain a fitting procedure involving five adjustment parameters, $a_1, a_2, a_3, \alpha$ and $\theta_D$.

Due to the very large unit cell 990.0(4) Å$^3$ [11], it is natural to expect rather low Debye temperatures for various acoustic phonon modes in MAPbI$_3$. Indeed, the results of the calorimetric study predict that the range of Debye temperatures is between 80 and 220 K for MAPbX$_3$ (X=Cl, Br, I) compounds [16]. Although for our SC samples some improvements of fits are obtained by using higher Debye temperatures $\theta_D$ in Eq. 1, hereafter we choose to fix $\theta_D$ at 120 K, in agreement with the value for the high acoustical mode reported previously[16] For the remaining four parameters our fit for SC samples provides the values displayed in the first row of Table 1, with a good quantitative agreement with experiments. Indeed, according to theoretical expectations, for the set of the SC parameters presented in Table 1, $\tau_u$ dominates the high-temperature behaviour, becoming negligible in the low-temperature limit $T \rightarrow 0$. The importance of the exponential factor in $\tau_u$, involving α, exhibits itself in the crossover region around $T_{max}$.

**Table 1. Fitting parameters by using Callaway's approach**

|  | $a_1/C$ | $a_2/C$ | $a_u/C$ | $\alpha$ |
|---|---|---|---|---|
| Single crystal | $2 \times 10^4$ | 0.004 | 0.87 | 16 |
| Polycrystal | $1.5 \times 10^5$ | 0.004 | 0.87 | 16 |

In order to extend our present approach to PC experimental data, we consider a particular fit keeping the relaxation rates $\tau_p^{-1}$ and $\tau_u^{-1}$ fixed to the values obtained for the SC sample, because these mechanisms should depend on the bulk properties. Despite of the fact that we are now dealing with just one free parameter, $a_1$, the fitting procedure yields a very satisfactory agreement with the experiment, comparable to the already discussed case of SC samples.

However, some aspects of the results in Table 1 deserve additional attention. First, $\kappa^{poly}$ near room temperature is reduced only by a factor of 2 compared to $\kappa^{single}$, which is one order of magnitude smaller than in other energy materials like CdSe [20]. Second, the effective cross section $S \propto \tau_b^2$ for all samples appears to be several orders of magnitude smaller than the specimen dimension. All this is a strong indication of a very effective, internal mechanism of acoustic phonon attenuation. Therefore, the values that follow from Table 1 for the cross section *S* should be understood as effective quantities, rather than being derived from the specimen dimension (the SC case), or from the grain boundaries (the PC case). In order to address this issue, we use a different mechanism, resonant scattering instead of scattering by impurities and grain boundaries. We keep the Umklapp scattering in the analysis because of its dominant contribution at high temperatures, $\tau_u^{-1} \propto T^3 x^2$. Resonant scattering is known to be of the great importance in materials with dynamical disorder, being frequently modelled [21] by $\tau_R^{-1} \propto a_r \frac{\omega_0^2 \omega^2}{(\omega_0^2 - \omega^2)^2}$. $\omega_0$ characterizes vibrations coupled to acoustic phonons. With $\theta_D = 120$ K, our fitting procedure involves three parameters: $a_u, a_R, \omega_0$. The computed curves nicely reproduce the experimental data, Figure 4, and the parameters of the model are summarized in Table 2.

**Table 2. Fitting parameters by assuming resonant and Umklapp scattering**

|  | $a_r/C$ | $\omega_0$ | $a_u/C$ |
|---|---|---|---|
| Single crystal | $4.8 \times 10^5$ | 42 K | 0.84 |
| Polycrystal | $5.8 \times 10^5$ | 42 K | 0.69 |

The obtained resonance frequency $\omega_0 = 42$ K in Table 2 may be conjectured to the slow, nondispersive phonon modes associated to the rotation degrees of freedom of $CH_3NH_3^+$. Indeed, rotational vibrations of this cation have recently been discussed [22], while for similar compounds, PbI based organic-inorganic hybrid, low energy optical modes were experimentally observed [23]. Furthermore, strong effects of the dynamical disorder on thermal conductivity were observed in other solids like clathrate hydrates and anion doped crystalline KCl [21, 24-25]. This leads us to a conclusion that the ultralow thermal conductivity in $CH_3NH_3PbI_3$ is due to its particular crystal structure, involving slowly rotating $CH_3NH_3^+$ cations within the unit cell.

In summary, we have presented thermal conductivity measurements for large single crystals and polycrystalline samples of $MAPbI_3$. The room temperature value of κ is equal to 0.5 W/(Km) for single crystals and 0.3 W/(Km) for polycrystals. These values are very low. The origin of the strongly reduced κ might be the disorder of the $CH_3NH_3^+$ sublattice and its easy excitation even below 160 K. Such a low κ will prevent the rapid spread of the light deposited heat, which can cause mechanical stresses and limit the lifetime of the photovoltaic device. These conclusions concerning κ stay valid even when the samples are exposed to light, since the number of photo-excited electrons remains low and their contribution to the thermal conductivity is negligible. Finally, the electrical and thermal conductivities in the 4–300 K temperature range clearly revealed the effect of grain boundaries but their importance is lower than expected.

**Experimental Methods**

Resistivity measurements on SC and PC samples were performed in a standard four-point configuration. Gold wires were glued on pre-evaporated gold pads on the sample. The experiments were carried out in the dark to avoid unwanted photo induced effects inside a closed-cycle He cryostat maintained at a base pressure of 1e-3mbar. Temperature was sweeped from 310 K down to 25 K. Thermal conductivity (κ) was measured by a steady-state method by using calibrated stainless-steel as reference sample (see fig 5). Special care was taken in maintaining temperature gradient across the sample around 1 K as described elsewhere [26].

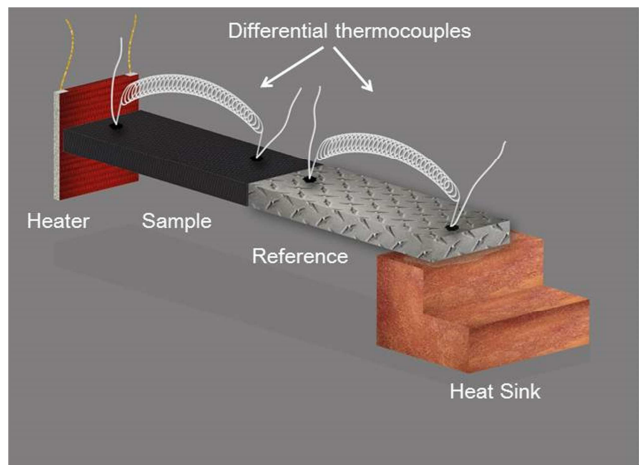

**Figure 5.** Schematic view of the experimental setup used for thermal conductivity measurement. The black rectangle represents the sample, the gray one is the calibrated stainless-steel reference and the orange solid is the copper heat sink. The gray curled wires are type E differential thermocouples connected to the sample and to the reference with Stycast 2850 Ft thermal conducting epoxy. A heater is connected to the sample using the same epoxy.

AUTHOR INFORMATION

**Corresponding Author**

*andrea.pisoni@epfl.ch
**Notes**

The authors declare no competing financial interests.

ACKNOWLEDGMENT

The work was supported by the Swiss National Science Foundation for fundamental research.

The authors wish to acknowledge and thank Andrijana Drobnjak for the precious help in images preparations.